\documentclass[conference]{IEEEtran}
\IEEEoverridecommandlockouts
\usepackage{cite}
\usepackage{amsmath,amssymb,amsfonts}
\usepackage{algorithmic}
\usepackage{graphicx}
\usepackage{textcomp}
\usepackage{xcolor}
\def\BibTeX{{\rm B\kern-.05em{\sc i\kern-.025em b}\kern-.08em
    T\kern-.1667em\lower.7ex\hbox{E}\kern-.125emX}}

\usepackage{enumitem}
\setlist[itemize]{nosep, leftmargin=*}

\usepackage{hyperref}
\usepackage{url}
\newcommand{\aaf}{\vspace*{-6pt}}
\newcommand{\af}{\vspace*{-3pt}}

\newcommand{\MH}[1]{\textcolor{black}{#1}}

\newcommand*\circled[1]{\tikz[baseline=(char.base)]{
            \node[shape=circle,draw,inner sep=0.6pt] (char) {#1};}}
\newcommand{\toolvyper}{{\texttt{Sol2Vy}}}

\usepackage[most,many]{tcolorbox}
\usepackage{tikz}
\usepackage{amsthm}
\usepackage{booktabs}
\usepackage{multirow}
\usepackage{subcaption}
\usepackage[normalem]{ulem}
\useunder{\uline}{\ul}{}
\usepackage{wrapfig}
\usepackage{makecell}
\usepackage{colortbl}
\definecolor{lightgray}{gray}{0.93}

\begin{document}


\title{Zero-Shot Vulnerability Detection in Low-Resource Smart Contracts  Through Solidity-Only Training}

\author{\IEEEauthorblockN{Minghao Hu}
\IEEEauthorblockA{\textit{Department of Computer Science} \\
\textit{George Mason University}\\
Fairfax, USA \\
mhu20@gmu.edu}
\and
\IEEEauthorblockN{Qiang Zeng}
\IEEEauthorblockA{\textit{Department of Computer Science} \\
\textit{George Mason University}\\
Fairfax, USA \\
zeng@gmu.edu}
\and
\IEEEauthorblockN{Lannan Luo}
\IEEEauthorblockA{\textit{Department of Computer Science} \\
\textit{George Mason University}\\
Fairfax, USA \\
lluo4@gmu.edu}
}

\maketitle

\begin{abstract}
Smart contracts have transformed decentralized finance, but flaws in their logic still create major security threats. Most existing vulnerability detection techniques focus on well-supported languages like Solidity, while low-resource counterparts such as Vyper remain largely underexplored due to scarce analysis tools and limited labeled datasets. Training a robust detection model directly on Vyper is particularly challenging, as collecting sufficiently large and diverse Vyper training datasets is difficult in practice.
To address this gap, we introduce \toolvyper, a novel framework that enables cross-language knowledge transfer from Solidity to Vyper, allowing vulnerability detection on Vyper using models trained exclusively on Solidity. 
This approach eliminates the need for extensive labeled Vyper datasets typically required to build a robust vulnerability detection model. We implement and evaluate \toolvyper~ on various critical vulnerability types, including reentrancy, weak randomness, and unchecked transfer. Experimental results show that \toolvyper, despite being trained exclusively on Solidity, achieves strong detection performance on Vyper contracts and significantly outperforms prior state-of-the-art methods.
\end{abstract}

\begin{IEEEkeywords}
smart contract, vulnerability detection, Solidity, Vyper
\end{IEEEkeywords}

\section{Introduction}

Smart contracts have become the foundation of decentralized applications and finance~\cite{25dsn-phishinghook,de2025scamdetect,liao2024bftrand}, yet their vulnerabilities continue to cause severe security and financial incidents. Detecting these vulnerabilities is therefore of paramount importance. Existing approaches fall into two categories. \emph{The first relies on traditional program analysis techniques}, such as Mythril~\cite{sharma2022survey}, Slither~\cite{feist2019slither}, and Securify~\cite{tsankov2018securify}. However, static analysis tools depend heavily on manually crafted patterns and often suffer from high false positive rates, especially on newer compiler versions, while fuzzing-based dynamic approaches struggle with path explosion, limiting their practical applicability. \emph{The second category leverages deep learning}~\cite{boi2024smart,liu2024propertygpt,sendner2023smarter,so2021smartest}, which offers greater scalability and accuracy by automatically learning semantic features from large datasets.

Existing deep learning-based approaches, however, overwhelmingly target Solidity, the dominant Ethereum smart contract language, which benefits from large, labeled datasets with well-curated vulnerability samples to support the training of robust detection models. In contrast, other low-resource languages such as Vyper remain severely underexplored. Vyper is a rising alternative with Python-like syntax and a security-oriented design that emphasizes simplicity, readability, and a reduced attack surface, making it attractive to developers who prioritize correctness and safety~\cite{Vyper}. It is well known that training deep learning models requires a large amount of labeled data. However, \emph{due to the data scarcity of labeled vulnerability samples in Vyper, no deep learning-based models currently exist for detecting vulnerabilities in Vyper}.

%

\vspace{3pt}
\noindent \textbf{Our Idea and Approach.} 
In this work, we take \emph{the first step toward vulnerability detection in Vyper smart contracts}. To address the challenge of data scarcity, we propose \emph{zero-shot vulnerability detection in Vyper via Solidity-only training}, where transferable knowledge is learned from Solidity, enabling a model trained on Solidity to detect vulnerabilities in Vyper. Achieving such cross-language knowledge transfer is non-trivial, however, due to the substantial syntactic and semantic differences between the two languages.

To learn transferable knowledge, we leverage SlithIR~\cite{feist2019slither}, a language-agnostic intermediate representation (IR) that serves as a bridge between Solidity and Vyper. SlithIR encodes smart contract semantics in a simplified, language-independent form, enabling effective cross-language representation learning. Building on this, we propose a novel three-stage framework, named \toolvyper, that systematically transfers vulnerability detection knowledge from Solidity to Vyper. 

Both Solidity and Vyper contracts are lifted to SlithIR using Slither. The \emph{first stage} is \emph{unsupervised transferable knowledge learning}, employs a multi-view architecture designed to capture both sequential and hierarchical representations of SlithIR. We train this architecture by minimizing the Maximum Mean Discrepancy (MMD) loss between Solidity and Vyper, enabling the model to learn \emph{transferable, language-agnostic knowledge} from unlabeled datasets. In the \emph{second stage}, \emph{supervised vulnerability detection on Solidity}, a classification network is added on top of the pre-trained multi-view architecture. This network is trained only using labeled vulnerable and safe \emph{Solidity contracts}, while the parameters of the multi-view backbone are frozen to preserve the transferable knowledge learned in the first stage. The \emph{third stage}, \emph{testing on Vyper}, applies the trained architecture and classification network directly to Vyper contracts without any modifications. Notably, \toolvyper\ achieves effective vulnerability detection in Vyper \emph{without requiring any labeled Vyper vulnerability samples for training}. This zero-shot capability addresses the data scarcity challenge in low-resource smart contract languages, overcoming both syntactic and semantic differences that have historically limited cross-language model reuse.

\vspace{3pt}
\noindent \textbf{Results.}
We implement our approach, \toolvyper, which learns transferable knowledge and enables effective vulnerability detection in Vyper contracts using models trained solely on Solidity. When the vulnerability detection model is trained and tested on Solidity, it achieves false positive rate (FPR) /false negative rate (FNR) scores of 0.09/0.13, 0.08/0.12, and 0.11/0.14 for the three vulnerability types: reentrancy (RE), weak randomness (WR), and unchecked transfer (UT), respectively.
When~\toolvyper\ is trained on Solidity and directly reused to evaluate Vyper contracts, it attains FPR/FNR scores of 0.11/0.16, 0.10/0.15, and 0.12/0.17 for RE, WR, and UT. The corresponding FPR/FNR increases (0.02/0.03, 0.02/0.03, and 0.01/0.03 for RE, WR, and UT) are minimal, indicating that \toolvyper\ successfully captures strong cross-language transferable knowledge, enabling a Solidity-trained model to perform reliably on Vyper. Moreover, our approach consistently outperforms baseline methods across all evaluated settings.
Below we highlight our contributions: 
\begin{itemize}
\item We take the first step toward enabling robust vulnerability detection for the low-resource Vyper language. \toolvyper\ performs zero-shot detection on Vyper contracts using only Solidity training data, removing the need for large labeled Vyper vulnerability datasets.


\item We design a novel three-stage pipeline that lifts Solidity and Vyper into SlithIR and learns language-agnostic semantic representations. A multi-view encoder captures both sequential instruction flows and hierarchical structures, providing richer and more comprehensive semantic coverage.

\item A classifier trained solely on labeled Solidity vulnerabilities generalizes effectively to Vyper, achieving low FPRs/FNRs (0.10–0.12) across RE, WR and UT. \toolvyper\ significantly outperforms prior state-of-the-art methods.

\item The Vyper vulnerability detector requires no labeled Vyper contracts, yet still accurately identifies vulnerable Vyper smart contracts, demonstrating the practical benefits of our cross-language transferable learning approach.

\end{itemize}

\section{Related Work}


\vspace{3pt}
\noindent \textbf{Ethereum Virtual Machine and Smart Contracts.}
Blockchain is a paradigm in distributed computing that underpins various cryptocurrency platforms and decentralized applications~\cite{ranchal2024zlb,perez2021smart,sharma2023mixed,chen2020understanding,albert2018ethir,yi2022empirical,jin2022exgen,ruggiero2024sok}. There are various blockchain platforms with unique features~\cite{polkadot,EOSIO,binance}. Ethereum Virtual Machine (EVM) pioneered programmable blockchains through its robust functionality~\cite{ma2023abusing,ma2023loki,jiao2020semantic,so2020verismart}. Smart contracts, written in various languages, are self-executing programs deployed on various platforms. Solidity~\cite{Solidity} remains the dominant language for EVM, while Vyper~\cite{Vyper} provides a more Pythonic alternative with enhanced security features. The variety of smart contract languages introduces fundamental distinctions in their security models, vulnerability patterns, and development paradigms, making cross-language analysis challenging. 



\vspace{3pt}
\noindent \textbf{Vulnerability Detection in Solidity Smart Contracts.}
Smart contract vulnerabilities pose significant financial risks to blockchain ecosystems, with millions of dollars lost due to code defects~\cite{smolka2023fuzz,cui2022vrust,liu2025anomaly,chen2025chatgpt,wu2024advscanner,luo2024scvhunter,zheng2023turn,bram2021rich,ma2024combining,grossman2024practical,wu2024we,wong2024confuzz,sharma2023mixed,perez2021smart}. Vulnerability detection in smart contracts has progressed through multiple approaches. Traditional static approaches include Slither~\cite{feist2019slither} and Smartcheck~\cite{tikhomirov2018smartcheck} employ various techniques to identify vulnerabilities without executing the code. Dynamic analysis has evolved through tools like Echidna~\cite{grieco2020echidna}, Mythril~\cite{sharma2022survey}, ContractFuzzer~\cite{jiang2018contractfuzzer}, symvalic~\cite{smaragdakis2021symbolic}, DivertScan~\cite{liu2025detecting}, ILF~\cite{he2019learning}, RLF~\cite{su2022effectively}, xFuzz~\cite{xue2022xfuzz}, CrossFuzz~\cite{yang2024crossfuzz}, sFuzz~\cite{nguyen2020sfuzz}, SMARTIAN~\cite{choi2021smartian}, SCFuzzer~\cite{wu2024we}, FunFuzz~\cite{ye2024funfuzz} and Harvey~\cite{wustholz2020harvey}. Recent work explores deep learning and LLMs for blockchain security analysis~\cite{boi2024smart,liu2024propertygpt,sendner2023smarter,so2021smartest,ijcai2021p379,hu2023large,chen2024improving,sun2024gptscan,cui2022vrust,smolka2023fuzz,zeng2025janusvln,zeng2025FSDrive,zhou2026risk,zhou2025beyond,xiaoyuema,li2025rowhammer,chen2025tracking,hu2025flowmaltrans,lei2026multiscalegraphlearningframework,shen2025aienhanced,shen2026mftformer,sun2025ft2,zhu2025understanding,sun2025demystifying,11426811,NEURIPS2023_944ecf65,shi2026codeocr,shi2026reasoning,chen2025progressive,xie2026rethinking,xie2024mmpalm,xie2024palm,zeng2026halluguard,zeng2025lensllm,guo2024take,he2023cross,10.1145/3785706.3785729,li2024synergized,Alghamdi2026Blockchain,Cao2025Privacy,lan2025contextual,lan2025mappo,li2026layertargetedmultilingualknowledgeerasure,codes2026,lai2026transformers,chen2025autoneural,cheng2026enhancingfinancialreportquestionanswering}. Clear~\cite{chen2024improving} captures the correlation between vulnerable and safe programs using a contrastive learning (CL) model. GPTLens~\cite{hu2023large} introduces two roles of detection: auditor and critic to enhance vulnerability discovery. AmeVulDetector~\cite{ijcai2021p379} combines vulnerability-specific patterns with neural networks. GPTScan~\cite{sun2024gptscan} combines LLM's understanding with static analysis checks by breaking vulnerabilities into scenarios and properties to query the LLM.

\vspace{3pt}
\noindent \textbf{Vulnerability Detection in Low-resource language.}
Despite the robust ecosystem of vulnerability detection tools for Solidity smart contracts, there remains a critical shortage of analysis capabilities for low-resource languages such as Vyper. For example, Slither and Mythril rely on manually crafted patterns or rules to detect vulnerabilities in Vyper, which has low coverage. Deep learning models often require large labeled datasets, which are scarce in low-resource languages like Vyper. This disparity in security tooling across smart contract languages creates a significant vulnerability gap in the blockchain ecosystem.

To mitigate this gap, we propose \emph{a novel framework that learns transferable, language-agnostic knowledge} across Solidity and Vyper, enabling the use of Solidity’s abundant labeled datasets to train a robust detection model that can be directly reused to analyze Vyper contracts

\section{Overview}


\subsection{Two Types of Datasets}

\toolvyper~aims to develop a robust deep learning model that can extend vulnerability detection capabilities from Solidity to Vyper. We define two types of datasets used in \toolvyper.

\vspace{3pt}
\noindent \textbf{General Dataset.}
This dataset is used to learn transferable language-agnostic knowledge shared between Solidity and Vyper. It consists of \emph{unlabeled} Solidity and Vyper smart contracts, which can be easily collected from public blockchain explorers (e.g., Etherscan) or open-source repositories. Since \emph{no annotation effort is required}, this unlabeled corpus enables our system to capture common semantic patterns, structural behaviors, and execution characteristics of smart contracts across the two languages, forming the foundation for effective cross-language transfer in later stages.

\vspace{3pt}
\noindent \textbf{Vulnerability Detection Dataset.} 
This dataset is related to the vulnerability detection task. It consists of both safe and vulnerable contracts in Solidity, and is used to train the vulnerability detection model. Our goal is to \emph{reuse} this Solidity-trained model to detect vulnerabilities in Vyper contracts by leveraging the transferable knowledge learned from the general unlabeled dataset.

When referring to data scarcity, we specifically mean the \emph{scarcity of Vyper vulnerability datasets}. While large numbers of \emph{unlabeled} contracts can be collected for the general dataset, and abundant labeled vulnerability samples exist for Solidity due to its maturity and dominance, \emph{Vyper lacks sufficiently labeled vulnerable contracts} because it is relatively new and supported by fewer security tools. As a result, directly training a robust vulnerability detector for Vyper is difficult. This motivates our approach of training the detector on Solidity and reusing it for Vyper vulnerability detection.


\subsection{IR Representation}

\begin{figure*}[]
  \centering
    \includegraphics[width=0.9\linewidth]{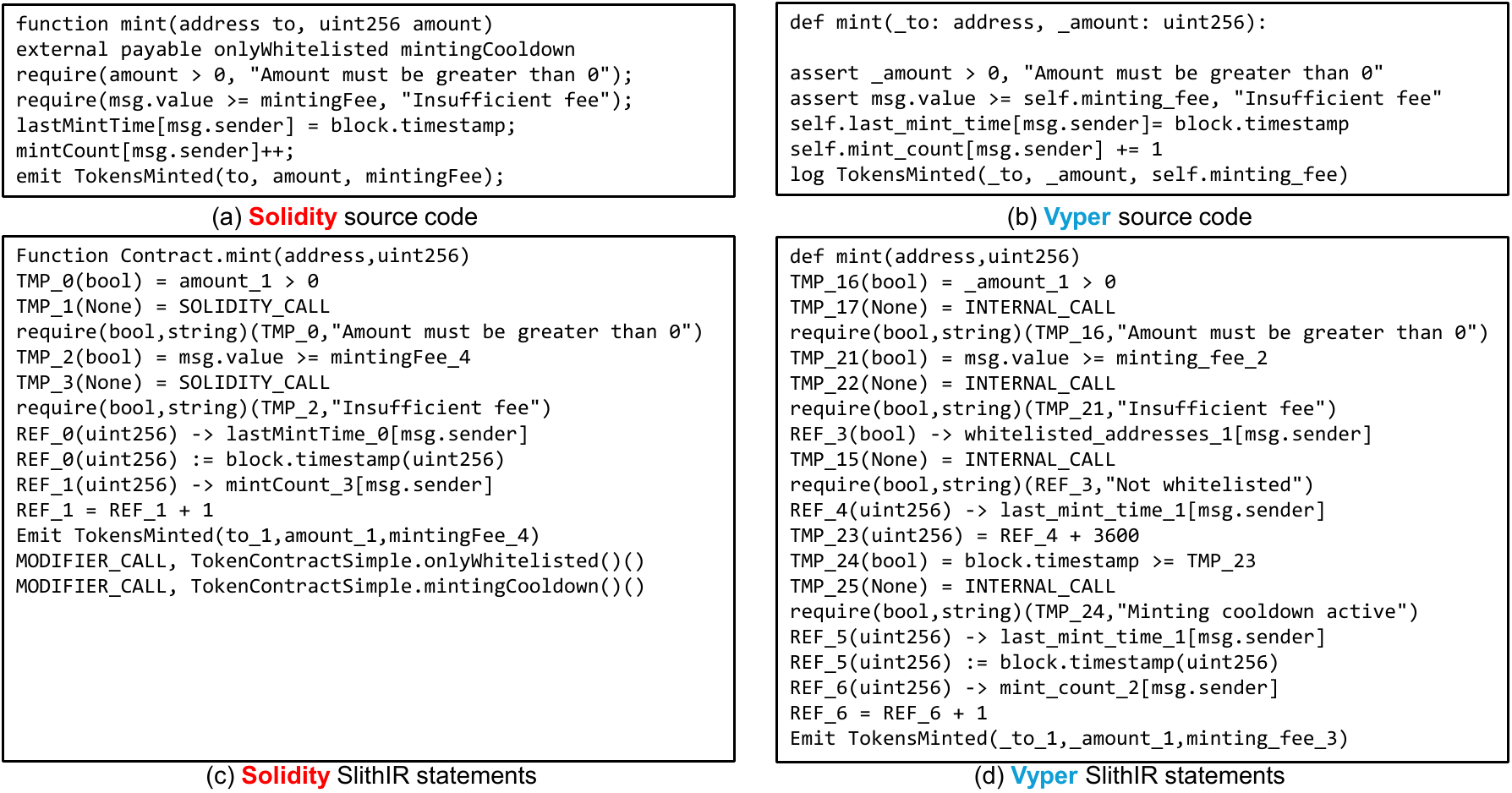}
    \af
  \caption{Comparison of SlithIR generated from Solidity and Vyper versions of the \texttt{mint()} function.}
  \label{fig:ir-process}
  \af
\end{figure*}

Solidity and Vyper express the same functionality using different source-level syntax. As shown in Figure \ref{fig:ir-process}(a) and Figure \ref{fig:ir-process}(b), even semantically equivalent implementations (e.g., the \texttt{mint()} function) differ substantially in structure, syntax, and language constructs. This motivates the need for a common representation that can abstract away language-specific syntax. To this end, we adopt SlithIR, a language-agnostic intermediate representation produced by the Slither~\cite{feist2019slither}. SlithIR converts both Solidity and Vyper into a standardized three-address code format that captures code core semantics.

However, \emph{directly relying on SlithIR is insufficient}. While SlithIR provides a unified format, the SlithIR generated from Solidity and Vyper can still differ significantly. Figures \ref{fig:ir-process}(c) and \ref{fig:ir-process}(d) illustrate that even when the source code implements identical functionality, their SlithIR forms remain noticeably different. These discrepancies arise from several factors, including: (1) \emph{Compilation-specific variations}: \texttt{solc} and \texttt{vyperlang} introduce distinct IR constructs (e.g., Solidity uses \texttt{MODIFIER\_CALL}, while Vyper inlines modifier-like logic); and (2) \emph{Differences in language features and control flow structures}: Given some unique high-level mechanism in Solidity (i.e., inheritance, overloading), the two languages generate divergent abstract syntax trees (AST) and thus different SlithIR, even for equivalent logic. As a result, in our experiment (Section~\ref{subsec:ablation}), a model trained solely on SlithIR derived from Solidity contracts performed poorly when tested on SlithIR derived from Vyper, confirming that SlithIR alone does not automatically align semantics across languages.

To address this misalignment, we design an \emph{unsupervised cross-language transfer learning pipeline} that learns a shared, language-agnostic representation of SlithIR (see Section~\ref{sec:ir}). Our multi-view encoder jointly models sequential instruction patterns and hierarchical structure, while an MMD-based alignment module minimizes the distributional distance between Solidity and Vyper SlithIR corpora. By training this encoder on large unlabeled datasets from both languages, we learn transferable semantic features that bridge the gap between their IRs.
These learned transferable representations form the foundation of our downstream vulnerability detector, enabling a classifier trained solely on Solidity vulnerabilities to effectively analyze Vyper contracts, without requiring any labeled Vyper vulnerability data.

\subsection{Model Architecture}

\begin{figure*}[t]
    \centering
    \includegraphics[width=0.82\linewidth]{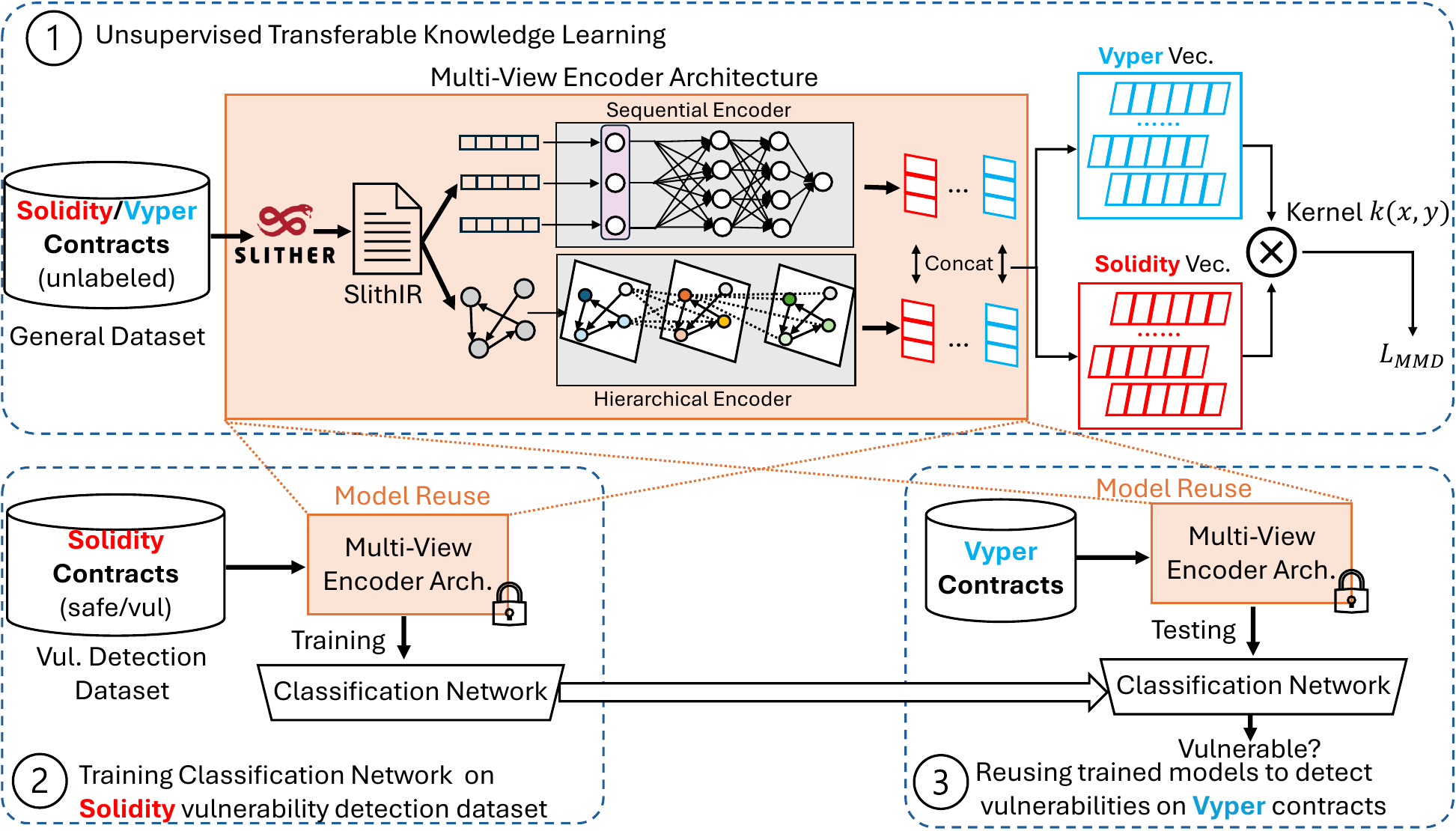}
    \aaf
    \caption{Applying~\toolvyper~to detect vulnerability in Vyper smart contracts by learning transferable knowledge from Solidity \&Vyper corpus and training a classification network on Solidity.}
    \label{fig:overview}
    
\end{figure*}

Our approach consists of three stages: (1) learning transferable knowledge in an unsupervised manner using the general dataset, (2) training a classification network on the Solidity vulnerability detection dataset, and (3) reusing the trained model to detect vulnerabilities in Vyper contracts.

Figure~\ref{fig:overview} shows an overview of~\toolvyper. In Step \circled{1}, we train the sequential and hierarchical encoders using the general dataset (consisting of \emph{unlabeled Solidity and Vyper contracts}) to learn transferable language-invariant knowledge. In Step \circled{2}, we train the vulnerability classification network using the Solidity vulnerability detection dataset, which contains \emph{labeled} safe and vulnerable \emph{Solidity} contracts. Finally, in Step \circled{3}, we directly reuse the trained encoders and classifier to test Vyper smart contracts, \emph{without any modification}. 

It is important to note that throughout the workflow, \toolvyper\ has never seen any vulnerable samples in Vyper, yet it still achieves strong vulnerability detection performance on Vyper contracts.


\section{Model Design}
\label{sec:ir}

\subsection{IR Preprocessing}

\vspace{3pt}
\noindent \textbf{IR Extraction.}
The first step is to lift both Solidity and Vyper source code into SlithIR. SlithIR abstracts away the syntactic differences of the two languages, providing a standardized, three-address code format that captures the core semantics and control flow. We first extract the ASTs through the compilers (\texttt{solc} for Solidity and \texttt{Vyperlang} for Vyper). Once we have the ASTs, we 
run Slither on the compiler outputs to generate per-function SlithIR, including (1) linearized IR statements, (2) basic blocks, and (3) call sites/event emissions/storage reads-writes. This produces a uniform IR interface for both languages even when their source constructs differ (e.g., modifiers, inheritance, and Vyper’s Pythonic syntax). We extract SlithIR at the function level (and keep function boundaries), because most vulnerability patterns (reentrancy, unchecked transfer, weak randomness) manifest within a function’s call/control/data interactions. Contract-level context (e.g., state variables) is retained through referenced storage symbols and types.

\vspace{3pt}
\noindent \textbf{IR Tokenization.}
We decompose each SlithIR instruction into constituent components based on operation types, operands, and control flow structures. The tokenization process transforms the three-address code format of SlithIR into discrete tokens while preserving semantic information critical for vulnerability detection: (1) Operation Tokenization: Core operations (e.g., \texttt{ASSIGN}, \texttt{CONDITION}) are preserved as atomic tokens to maintain instruction semantics. (2) Variable Tokenization: State variables, local variables, and temporary variables (e.g., \texttt{TMP\_0}) are tokenized separately to distinguish their scope and lifetime. (3) Type Information: Type annotations (e.g., \texttt{uint256}, \texttt{bool}, \texttt{address}) are retained as separate tokens to preserve type-safety semantics. 
(4) Control Flow Markers: Branch conditions, function boundaries, and expression hierarchies are marked with splitters to maintain structural information. 

\vspace{3pt}
\noindent \textbf{IR Normalization.}
After tokenization, we process SlithIR to reduce compilation-specific noise. Specifically, the following normalization rules are applied: 

\begin{itemize}
    \item The names of temporary variables, reference variables and tuple variables are normalized: \texttt{TMP\_0} $\rightarrow$ \texttt{<TMP>}, \texttt{REF\_3} $\rightarrow$ \texttt{<REF>}, \texttt{TUP\_2} $\rightarrow$ \texttt{<TUP>}.
    \item Simplify the customized variable name: \texttt{amount\_1} $\rightarrow$ \texttt{amount}, \texttt{rate\_2} $\rightarrow$ \texttt{rate}.
    \item We remove messages from statements: 
     \texttt{require (<TMP>,"Amount > 0")} $\rightarrow$ \texttt{require (<TMP>)}.
    \item The string literals  are replaced with \texttt{<STR>}. 
\end{itemize}

\subsection{Learning Transferable Knowledge}

\noindent \textbf{Multi-View Architecture.}
To capture the sequential and structural semantic information inside SlithIR, we design a multi-view architecture to encode \emph{both the sequential and structural} architecture of SlithIR. The multi-view architecture consists of two components that operate in parallel on SlithIR:

The first component is a \textbf{\em sequential encoder}, which processes SlithIR instructions as a sequence to capture temporal dependencies and execution flow patterns.  
We employ a Transformer-based architecture for the sequential encoder, leveraging its proven effectiveness in modeling long-range dependencies in sequential data. It consists of the following parts:
(1) Token embedding layer: The tokenized SlithIR instructions are first converted into dense vector representations. Each token (operation, variable, type annotation) is mapped to an embedding vector via a learnable embedding matrix $E \in R^{|V|\times d_{\text{seq}}}$, where $|V|$ is the vocabulary size and $d_{\text{seq}}$ is the embedding dimension (we set it to 128 in implementation). 
(2) Multi-head self-attention mechanism: The core of our sequential encoder employs multi-head self-attention to capture dependencies between the instructions regardless of their distance in the sequence. This allows the model to jointly attend to information from different representation subspaces, capturing diverse semantic relationships such as data flow dependencies, control flow patterns and variable usage patterns.
(3) Feed-forward networks: Each Transformer layer includes a position-wise feed-forward network consisting of two linear transformations with ReLU activation.


The second component is a \textbf{\em hierarchical encoder}, which processes the Abstract Syntax Tree (AST) derived from SlithIR to capture structural relationships and architectural patterns. 
We construct a graph representation $G=(V,E)$ where nodes $V$ correspond to syntactic elements such as functions, control flow constructs, variable declarations, and expression components. The edges $E$ in this graph represent three types of relationships: (1) AST structural relationships $E_{\text{ast}}$, (2) Variable definition-use chains $E_{\text{data}}$ and (3) Execution flow between basic blocks $E_{\text{ctrl}}$. 

To process this graph structure, we employ Graph Attention Networks (GATs), which provide the capability to learn adaptive attention weights for different types of structural relationships. For node feature initialization, each node $v \in V$ is initialized with a feature vector $x_v \in R^{d_{\text{hie}}}$ (where we use $d_{\text{hie}}=128$ in our implementation) that encodes: (1) Node type embedding, (2) Operation type and (3) Variable scope information (state/local/temporary). 

After node initialization, the GAT layers aggregate information from neighboring nodes through learned attention weights, enabling the propagation of contextual information throughout the hierarchical structure. This aggregation process captures multi-hop relationships in the graph, allowing the model to understand complex structural dependencies that may span multiple levels of the syntax tree. The hierarchical encoder thus produces node-level representations that encode both local structural properties and global architectural patterns.



\vspace{3pt} \noindent \textbf{Feature Extraction and Combination.}
For each Solidity and Vyper smart contract in the general dataset, we lift it to SlithIR and preprocess it. Then we feed the processed SlithIR through our multi-view architecture. The sequential encoder processes the tokenized SlithIR instructions and outputs a feature vector $h_{\text{seq}} \in \mathbb{R}^{d_{\text{seq}}}$. The hierarchical encoder processes the AST-derived graph structure and outputs a feature vector $h_{\text{hie}} \in \mathbb{R}^{d_{\text{hie}}}$. We concatenate these presentations to form the final feature vector: $z=[h_{\text{seq}}, h_{\text{hie}}]$ (see step \circled{1} in Figure~\ref{fig:overview}, where $d=d_{\text{seq}}+d_{\text{hie}}$ is the total dimensionality of the combined feature representation. We set both $d_{\text{hie}}$ and $d_{\text{seq}}$ to 128 so the total dimensionality is $d=256$. 

\vspace{3pt} \noindent \textbf{Maximum Mean Discrepancy (MMD) Loss.}
The fundamental challenge 
is the domain gap of the SlithIR code between Solidity and Vyper. Even though they both follow a similar format, their SlithIR representations exhibit distributional differences due to compiler-specific optimizations, language-specific idioms, and syntactic variations.


To address this challenge, we need a learning objective that explicitly encourages the encoder to learn representations that are \emph{invariant to the source language} while preserving the semantic information necessary for vulnerability detection. This is the role of Maximum Mean Discrepancy (MMD) loss in our framework. We employ MMD to minimize the distributional distance between Solidity and Vyper feature representations. MMD is a non-parametric statistical method that measures the distance between two probability distributions by comparing their mean embeddings in a Reproducing Kernel Hilbert Space (RKHS). Specifically, given a batch of feature vectors from  Solidity ($z_1^S, z_2^S, \dots,z_{n_S}^S $) and Vyper ($z_1^V, z_2^V, \dots,z_{n_V}^V $). The MMD loss $\mathcal{L}_{\text{MMD}}$ is calculated in Equation~(\ref{equ:mmd}).

\aaf
\begin{equation}
\label{equ:mmd}
\small
\mathcal{L}_{\text{MMD}} = \left\| \frac{1}{n_S} \sum_{i=1}^{n_S} \phi(\mathbf{z}_i^S) - \frac{1}{n_V} \sum_{j=1}^{n_V} \phi(\mathbf{z}_j^V) \right\|_{\mathcal{H}}^2
\end{equation}

where $\phi(\cdot):\mathbb{R}^d\rightarrow\mathcal{H}$ is the feature mapping function that maps input features into the RKHS $\mathcal{H}$, 

\vspace{3pt}\noindent \textbf{Kernel Design and Implementation.}
The feature mapping function $\phi(\cdot)$ is implicitly defined through the kernel trick using a combination of linear and Gaussian Radial Basis Function (RBF) kernels. We use a composite kernel defined in Equation~(\ref{equ:kernel}).
\aaf 
\begin{equation}
\label{equ:kernel}
\small
k(x,y) = \alpha k_{\text{Linear}}(x,y) + (1-\alpha) k_{\text{RBF}}(x,y)
\end{equation}

where $k_{\text{Linear}}(x,y)=x^Ty$ captures linear relationships between features. And $k_{\text{RBF}}(x,y)=\exp(-\gamma\|(x-y)\|^2)$ captures non-linear similarities. $\alpha$ is the bandwidth hyperparameter that balances the contribution of linear and non-linear components. In practice, we use Equation~(\ref{equ:mmd-implicit}) to calculate $\mathcal{L}_{\text{MMD}}$ to avoid explicit computation of the feature mapping $\phi(\cdot)$ while still measuring the distance between language distributions in RKHS $\mathcal{H}$. This transferable knowledge learning stage operates in a fully unsupervised manner, \emph{requiring no labeled Vyper data}. This aligns perfectly with our problem setting, where labeled Vyper datasets are scarce. The loss only requires unlabeled samples from both domains, making it practical for low-resource language scenarios.
\aaf
\begin{equation}
\small
\label{equ:mmd-implicit}
\begin{split}
    \mathcal{L}_{\text{MMD}} 
    &= \frac{1}{n_S^2} \sum_{i=1}^{n_S} \sum_{i'=1}^{n_S} k(\mathbf{z}_i^S, \mathbf{z}_{i'}^S) 
    + \frac{1}{n_V^2} \sum_{j=1}^{n_V} \sum_{j'=1}^{n_V} k(\mathbf{z}_j^V, \mathbf{z}_{j'}^V) \\
    &\quad - \frac{2}{n_S n_V} \sum_{i=1}^{n_S} \sum_{j=1}^{n_V} k(\mathbf{z}_i^S, \mathbf{z}_j^V)
\end{split}
\end{equation}

\subsection{Vulnerability Detection Module}

Once the multi-view encoders are trained to produce language-agnostic representations, we proceed to the vulnerability detection task, leveraging the learned transferable knowledge. This process involves the following stages: supervised training on Solidity, and cross-language testing on Vyper.

\vspace{3pt}
\noindent \textbf{Training the Classification Network Using \emph{Solidity} Vulnerability Detection Dataset.} In this stage, the weights of the sequential and hierarchical encoders from Stage 
\circled{1} are frozen to preserve the learned transferable knowledge. This ensures that the vulnerability-specific patterns learned from Solidity do not corrupt the language-agnostic representations. A classification network is added on top of the frozen encoders for vulnerability detection. We train the network using labeled Solidity smart contracts (safe and vulnerable), handling each vulnerability type separately.

The classification network consists of a multi-layer perceptron (MLP). It takes the concatenated feature vector $z=[h_{\text{seq}}, h_{\text{hie}}] \in \mathbb{R}^d$ from the frozen encoders as input, where $d=d_{\text{seq}}+d_{\text{hie}}$ represents the combined dimensionality. We employ two fully connected hidden layers with ReLU activation functions. The final output layer produces vulnerability predictions using a sigmoid function for binary classification (vulnerable or safe). For each vulnerability type (RE, WR, UT), we train a separate classification network.

\vspace{3pt}
\noindent \textbf{Reusing the Trained Model to Test on Vyper.} After training, the model can be applied directly to test Vyper smart contracts. The multi-view encoders and classification network process each contract to predict whether it is safe or vulnerable, without any additional fine-tuning or modification.

\section{Evaluation}

We conduct extensive experiments to assess \toolvyper\ in terms of vulnerability detection effectiveness. Additionally, we perform hyperparameter studies, runtime and few-shot analysis. We address seven research questions (\textbf{RQs}):

\begin{tcolorbox}[
  title=Research Questions,
  coltitle=white,
  colback=white,
  colframe=black,
  colbacktitle=gray!60!black,
  fonttitle=\bfseries,
  boxrule=0.5pt,
  arc=6pt,
  width=\linewidth,
  enhanced,
  borderline={0.5pt}{0pt}{dashed}
]
\aaf
\textbf{RQ1:} How effective is the language-agnostic features alignment for learning transferable knowledge?

\textbf{RQ2:} What is the vulnerability detection performance of~\toolvyper?

\textbf{RQ3:} How does~\toolvyper~ compare with baseline methods?

\textbf{RQ4:} How do different hyperparameter settings impact vulnerability detection performance?

\textbf{RQ5:} How does each component in~\toolvyper\ affect the overall vulnerability detection performance?

\textbf{RQ6:} What are the effects of adding few-shot Vyper samples in Stage \circled{2}?

\textbf{RQ7:} What is the runtime overhead for~\toolvyper? 


\aaf
\end{tcolorbox}



\subsection{Experimental Setting}

We implement \toolvyper\ using Transformer and Graph Attention Network in Pytorch 2.0. The sequential encoder is built with a 6-layer Transformer architecture with 8 attention heads and 128-dimensional embeddings, while the hierarchical encoder employs a 3-layer Graph Attention Network (GAT) with 128-dimensional node features. The classification network consists of a 2-layer MLP with ReLU activation and dropout ($p=0.3$). For MMD computation, we use a composite kernel with $\alpha=0.8$ and $\gamma=0.005$. All the experiments were conducted on a computer with Intel Core i9-12900K CPU (16 cores, 3.2GHz), 64GB RAM, and an NVIDIA RTX 4090 GPU with 24GB VRAM.

\vspace{3pt}
\noindent \textbf{General Dataset.}
We collect unlabeled Solidity and Vyper smart contracts from diverse and authoritative sources, including Etherscan~\cite{Etherscan} and GitHub repositories~\cite{github}. We collect $1842$ Solidity contracts and $1193$ Vyper contracts. 
It is important to note that the general dataset used to learn transferable language-agnostic knowledge has \emph{no overlap} with the vulnerability detection dataset used in the vulnerability detection task. 

\vspace{3pt}
\noindent \textbf{Learning Transferable Knowledge.}
We use the general dataset to train the multi-view architecture to learn language-invariant transferable knowledge.
We train the sequential and hierarchical encoder by minimizing the MMD loss between Solidity and Vyper. We employ a curriculum learning strategy where we gradually increase the complexity of the unlabeled contract dataset throughout training. We measure the complexity based on a combination of AST depth, SlithIR instruction diversity and contract scale. This approach helps the model learn fundamental transferable knowledge before tackling more challenging distributional differences. The training stops until the validation MMD loss drops below $0.2$.

\subsection{RQ1: Evaluating Quality of Transferable Knowledge}
\label{sec:rq1}


To validate our assumption that MMD-based alignment captures meaningful semantic correspondence, we conduct a quantitative analysis of the learned feature representations. Specifically, our goals are threefold: (\textbf{G1}) Semantically equivalent Solidity-Vyper pairs have a high similarity; (\textbf{G2}) The learned features are truly language-agnostic; (\textbf{G3}) Prevent representation collapse, where the encoder maps all inputs to nearly identical vectors to trivially minimize MMD loss.  


\vspace{3pt}
\noindent \textbf{G1: Semantic Equivalency.}
We construct a set of semantically equivalent SlithIR statement pairs from Solidity and Vyper, respectively. The pairs are chosen from six representative patterns: balance update, authorization check, external call, array push, reentrancy pattern and randomness condition. For each pattern, we select the most frequent statement pairs for assessment. For example, \texttt{<REF>->balances[sender]} and \texttt{<REF>->self.balances[sender]} are an equivalent pair between Solidity and Vyper SlithIR that belongs to the balance update category. Table~\ref{tab:cross-lang-sim} shows that for all six representative patterns, the cosine similarity scores are significantly higher after MMD training, indicating that semantically equivalent statements are closer in feature space, regardless of their source language (\textbf{Achieving G1}).

\begin{table}[t]
\caption{Semantic similarity of equivalent SlithIR patterns.}
\aaf
\label{tab:cross-lang-sim}
\centering
\resizebox{0.8\columnwidth}{!}{
\begin{tabular}{c|c|c}
\hline
\begin{tabular}[c]{@{}c@{}}Pattern\\ Type\end{tabular} & \begin{tabular}[c]{@{}c@{}}Cos. Sim. \\ (Before)\end{tabular} & \begin{tabular}[c]{@{}c@{}}Cos. Sim. \\ (After)\end{tabular} \\ \hline \hline
Balance Update                                         &0.21&\textbf{0.42}\\
Authorization Check                                    &0.12&\textbf{0.39}\\
External Call                                          &0.13&\textbf{0.40}\\
Array Push                                             &0.15&\textbf{0.37}\\
Reentrancy Pattern                                     &0.18&\textbf{0.43}\\
Randomness Condition                                   &0.17&\textbf{0.39}\\ \hline
\end{tabular}
}\aaf
\end{table}

\vspace{3pt}
\noindent \textbf{G2: Language Discriminability.}
To quantitatively measure whether the features are truly language-agnostic, we train a simple logistic regression classifier to predict SlithIR statements from Solidity or Vyper. Before MMD training, this classifier achieves 94.3\% accuracy, indicating that features contain strong language-specific signals. After MMD training, classification accuracy drops to 52.1\% (near random chance), confirming that language-specific patterns have been effectively suppressed (\textbf{Achieving G2}).

\vspace{3pt}
\noindent \textbf{G3: Representation Collapse Prevention.} 
A critical concern in domain alignment methods is representation collapse, where the encoder might learn trivial solutions by mapping all inputs to identical or near-identical representations. To verify that our representations remain expressive, we measure the variance within each language's feature distribution. Results show that our intra-class variance remains stable, decreasing only slightly from 32.1 to 28.9 for Solidity and 39.5 to 37.6 for Vyper. This modest reduction indicates that MMD training compresses language-specific noise while preserving semantic diversity (\textbf{Achieving G3}). Representation collapse is prevented through two complementary mechanisms inherent to our framework design:
\begin{itemize}
    \item First, our multi-view architecture provides implicit regularization against collapse. The self-attention mechanisms within the sequential encoder preserve positional and ordering information that could be lost in collapsed representations. Similarly, the hierarchical encoder processes graph structures through GAT layers that aggregate neighborhood information, which requires diverse node representations. If representations collapsed, the attention weights in both encoders would become uninformative, preventing the model from learning useful patterns in the unlabeled corpora. 
    \item Second, we employ \textbf{\emph{curriculum learning}} during Stage \circled{1}, gradually increasing dataset complexity. This progressive training strategy encourages the model to first learn coarse-grained structural patterns before refining fine-grained semantic distinctions, naturally preventing premature convergence to degenerate solutions. 
\end{itemize}

\vspace{3pt} 
In summary, semantically equivalent Solidity-Vyper statements are mapped to a closer region in the feature space (\textbf{G1}). After MMD-based training, the encoders largely encode language-agnostic artifacts (\textbf{G2}). Furthermore, this alignment is achieved without sacrificing representational expressiveness (\textbf{G3}). These results collectively verify that our MMD-based multi-view alignment module \textbf{\emph{learns language-agnostic features}} that form a reliable foundation for zero-shot vulnerability detection on Vyper contracts.

\subsection{RQ2: Vulnerability Detection}
\label{sec:rq2}

After training the multi-view model (consisting of the sequential and hierarchical encoders) on the general datasets to learn transferable knowledge, we then use the \emph{Solidity} vulnerability detection dataset to train the classification network. Once trained, we directly reuse the encoders together with the Solidity-trained classifier to evaluate Vyper smart contracts, enabling cross-language vulnerability detection \emph{without requiring any Vyper-labeled data}.

\vspace{3pt}
\noindent \textbf{Vulnerability Detection Dataset in \emph{Solidity}.}
We build this dataset from multiple popular smart contract repositories like SmartBugs~\cite{smartbugs}, SWC-registry~\cite{swcregistry}, DAppSCAN~\cite{dapscan}, Etherscan~\cite{Etherscan}. Since existing datasets often contain inconsistent or incorrect annotations, we implement a systematic label verification process to ensure high data quality. For initial vulnerability labeling, we employ multiple established tools, including Slither~\cite{feist2019slither}, Securify~\cite{tsankov2018securify}, Mythril~\cite{sharma2022survey}, Echidna~\cite{grieco2020echidna}, which provide automated detection capabilities for known vulnerability patterns. We apply a consensus criterion where vulnerabilities must be identified by at least two independent tools to go through the subsequent manual verification. 

Then, we manually verify every candidate Solidity label with a two-reviewer, auditable checklist: for collected samples, a contract is first flagged only when at least two tools agree on the same vulnerability type, after which two reviewers independently confirm the label by recording (1) the vulnerability type (RE/WR/UT), (2) exact evidence locations (function + line range / IR snippet), and (3) a brief exploit rationale. RE requires identifying an external call enabling re-entry before critical state update, WR requires a security-relevant decision using manipulable entropy (e.g., timestamp/block data), and UT requires a value-transfer whose success is not checked or enforced by revert/require. Labels are accepted only when both reviewers agree on type and evidence. Otherwise, the sample is marked disputed and excluded.

After the process, we collect 1275, 753, 785 and 813 Solidity contracts that are labeled as safe, reentrancy (RE), weak randomness (WR) and unchecked transfer (UT), respectively. We allocate 80\% of the data for training (denoted as $D_{\text{Sol-Train}}$) and 20\% for testing (denoted as $D_{\text{Sol-Test}}$). Note that the vulnerability detection dataset \emph{has no overlap} with the general dataset used for learning transferable knowledge.

\vspace{3pt}
\noindent \textbf{Training Classification Layer on Solidity.}
We freeze the sequential and structural encoders and append a classification layer to the combined output. The classification network is trained on $D_{\text{Sol-Train}}$ until the loss falls below 0.5.

\vspace{3pt}
\noindent \textbf{Vulnerability Detection Dataset in \emph{Vyper}.}
Due to the limited availability of existing labeled Vyper datasets, we construct a vulnerability detection dataset for \emph{Vyper} from scratch in two ways. 
First, we collect Vyper contracts from public repositories and apply Slither~\cite{feist2019slither} and Mythril~\cite{sharma2022survey} (both of which support Vyper) to label them.  
Second, we create additional vulnerable contract variants by randomly introducing the three target vulnerability patterns into safe contracts. \MH{We go through the same labeling and manual verification process as the Solidity dataset. For injected samples, reviewers additionally confirm the injected code is reachable and semantically consistent (not dead/unexecutable) before inclusion.}

\begin{table}[t]
\centering
\caption{Vyper contract length (Line of Code, LoC).}
\aaf 
\label{tab:vyper-stat}
\resizebox{\columnwidth}{!}{
\begin{tabular}{ccccc}
\toprule
\MH{Type} & \MH{Total} & \MH{20--100 LoC} & \MH{101--200 LoC} & \MH{$\ge$201 LoC} \\
\midrule
\MH{Safe} & \MH{434} & \MH{120} & \MH{140} & \MH{174} \\
\MH{RE}   & \MH{236} & \MH{60}  & \MH{80}  & \MH{96}  \\
\MH{WR}   & \MH{226} & \MH{55}  & \MH{75}  & \MH{96}  \\
\MH{UT}   & \MH{307} & \MH{70}  & \MH{100} & \MH{137} \\
\midrule
\MH{All}  & \MH{1203} & \MH{305} & \MH{395} & \MH{503} \\
\bottomrule
\end{tabular}
}

\end{table}

Our final dataset comprises 434, 236, 226 and 307 Vyper contracts that are labeled as safe, RE, WR and UT, respectively. \MH{The details are shown in Table~\ref{tab:vyper-stat}.} All the safe contracts and 160, 143, and 205 RE, WR and UT vulnerable contracts are collected from public repositories, despite our significant efforts to gather Vyper vulnerable samples.

\begin{table*}[t]
\centering
\caption{A Solidity-trained model evaluated on both Solidity and Vyper samples.}
\label{tab:rq2}
\aaf
\resizebox{0.88\textwidth}{!}{
\begin{tabular}{ll|cc|cc|cc}
\hline
\multicolumn{2}{l|}{\multirow{2}{*}{}}                                  & \multicolumn{2}{c|}{RE}          & \multicolumn{2}{c|}{WR}          & \multicolumn{2}{c}{UT}           \\ \cline{3-8} 
\multicolumn{2}{l|}{}                                                   & FPR                       & FNR  & FPR                       & FNR  & FPR                       & FNR  \\ \hline \hline
\multicolumn{2}{l|}{Train and Test on Solidity (Optimal Setting)}       & \multicolumn{1}{c|}{0.09} & 0.13 & \multicolumn{1}{c|}{0.08} & 0.12 & \multicolumn{1}{c|}{0.11} & 0.14 \\ \hline
\multicolumn{2}{l|}{Train on Solidity \& Test on Vyper} & \multicolumn{1}{c|}{{0.13}} & {0.15} & \multicolumn{1}{c|}{{0.12}} & {0.14} & \multicolumn{1}{c|}{{0.13}} & {0.15} \\ \hline
\end{tabular}
}

\end{table*}

\begin{table*}[ht]
\centering
\caption{\MH{Comparison between~\toolvyper\ and Baseline Methods.}}
\aaf
\label{tab:rq3}
\resizebox{0.88\textwidth}{!}{
\begin{tabular}{lc|cc|cc|cc}
\hline
\multicolumn{2}{l|}{\multirow{2}{*}{}}                                                                                                                                                                          & \multicolumn{2}{c|}{RE}                               & \multicolumn{2}{c|}{WR}                               & \multicolumn{2}{c}{UT}                               \\ \cline{3-8} 
\multicolumn{2}{l|}{}                                                                                                                                                                                           & FPR                       & FNR                       & FPR                       & FNR                       & FPR                       & FNR                      \\ \hline \hline
\multicolumn{1}{l|}{\multirow{2}{*}{Program Analysis-based}}                                                                  & Slither                                                                         & \multicolumn{1}{c|}{0.46} & 0.58                      & \multicolumn{1}{c|}{0.48} & 0.55                      & \multicolumn{1}{c|}{0.45} & 0.52                     \\ \cline{2-8} 
\multicolumn{1}{l|}{}                                                                                                         & Mythril                                                                         & \multicolumn{1}{c|}{0.38} & 0.49                      & \multicolumn{1}{c|}{0.40} & 0.46                      & \multicolumn{1}{c|}{0.42} & 0.48                     \\ \hline
\multicolumn{1}{l|}{\multirow{3}{*}{\begin{tabular}[c]{@{}l@{}}Deep Learning-based \\ (Train \& Test on Vyper)\end{tabular}}} & \begin{tabular}[c]{@{}c@{}}Vyper-trained \\ Classification Network\end{tabular} & \multicolumn{1}{c|}{0.24} & 0.31                      & \multicolumn{1}{c|}{0.29} & 0.35                      & \multicolumn{1}{c|}{0.26} & 0.32                     \\ \cline{2-8} 
\multicolumn{1}{l|}{}                                                                                                         & CodeT5+                                                                         & \multicolumn{1}{c|}{0.28} & 0.34                      & \multicolumn{1}{c|}{0.32} & 0.38                      & \multicolumn{1}{c|}{0.30} & 0.36                     \\ \cline{2-8} 
\multicolumn{1}{l|}{}                                                                                                         & GraphCodeBERT                                                                   & \multicolumn{1}{l|}{0.32} & \multicolumn{1}{l|}{0.38} & \multicolumn{1}{l|}{0.35} & \multicolumn{1}{l|}{0.42} & \multicolumn{1}{l|}{0.33} & \multicolumn{1}{l}{0.39} \\ \hline

\multicolumn{1}{l|}{\multirow{3}{*}{\MH{LLM-based Methods}}}                                                                        & \MH{GPT-4.1}                                                                         & \multicolumn{1}{c|}{\MH{0.41}} & \MH{0.52}                      & \multicolumn{1}{c|}{\MH{0.44}} & \MH{0.56}                      & \multicolumn{1}{c|}{\MH{0.40}} & \MH{0.50}                     \\ \cline{2-8} 
\multicolumn{1}{l|}{}                                                                                                         & \MH{CodeLlama-7b}                                                                    & \multicolumn{1}{c|}{\MH{0.45}} & \MH{0.55}                      & \multicolumn{1}{c|}{\MH{0.47}} & \MH{0.59}                      & \multicolumn{1}{c|}{\MH{0.43}} & \MH{0.53}                     \\ \cline{2-8} 
\multicolumn{1}{l|}{}                                                                                                         & \MH{Qwen-Coder-7b}                                                                   & \multicolumn{1}{c|}{\MH{0.39}} & \MH{0.50}                      & \multicolumn{1}{c|}{\MH{0.42}} & \MH{0.54}                      & \multicolumn{1}{c|}{\MH{0.38}} & \MH{0.49}                     \\ \hline

\multicolumn{1}{l|}{\MH{Joint Training Baseline}}                                                                                             & \MH{\toolvyper~w/o freezing}                                                                       & \multicolumn{1}{l|}{\MH{0.48}} & \multicolumn{1}{l|}{\MH{0.53}} & \multicolumn{1}{l|}{\MH{0.68}} & \multicolumn{1}{l|}{\MH{0.78}} & \multicolumn{1}{l|}{\MH{0.47}} & \multicolumn{1}{l}{\MH{0.56}} \\ \hline

\multicolumn{1}{l|}{Our Approach}                                                                                             & \toolvyper                                                                       & \multicolumn{1}{l|}{\textbf{0.11}} & \multicolumn{1}{l|}{\textbf{0.16}} & \multicolumn{1}{l|}{\textbf{0.10}} & \multicolumn{1}{l|}{\textbf{0.15}} & \multicolumn{1}{l|}{\textbf{0.12}} & \multicolumn{1}{l}{\textbf{0.17}} \\ \hline
\end{tabular}
}
\end{table*}

\vspace{3pt}
\noindent \textbf{Results.}
We first evaluate the Solidity-trained classifier on the Solidity test dataset, $D_{Sol-test}$, which represents \emph{the upper bound of achievable performance} since the classifier is trained and tested on the same language. In this setting, no cross-language knowledge transfer is involved, and thus no inter-language information loss occurs. The classifier, trained and evaluated exclusively on Solidity contracts, \MH{achieves FPR/FNR values of 0.09/0.13, 0.08/0.12 and 0.11/0.14 for RE, WR and UT, respectively.}

We next reuse the Solidity-trained classifier to evaluate Vyper contracts using $D_{Vy-all}$. The results are shown in Table~\ref{tab:rq2}. It \MH{achieves FPR/FNR scores of 0.13/0.15, 0.12/0.14 and 0.13/0.15 for RE, WR and UT.} Note that this evaluation is performed in a zero-shot setting; \MH{when few-shot training with a balanced combination is applied, both FPR and FNR decrease substantially} (see Section~\ref{subsec:few-shot}).

Comparing the performance on Solidity and Vyper, we observe \emph{only minimal degradation}: an increase of just \MH{0.04/0.02, 0.04/0.02, and 0.02/0.01 in FPR/FNR} for RE, WR, and UT, respectively. This minimal degradation demonstrates the high quality of the transferable knowledge learned and the strong semantic preservation achieved by our approach.

To further understand why our model maintains strong performance across languages, we analyze how 
\toolvyper\ leverages transferable knowledge during vulnerability detection. 
\toolvyper\ effectively connects the learned cross-language semantic representations to the three vulnerability types (RE, WR, and UT) through pattern-type alignment.
As shown in Table~\ref{tab:cross-lang-sim} in RQ1 (\ref{sec:rq1}), the model aligns critical vulnerability-specific patterns such as ``Balance Update'' for RE and ``Authorization Check'' for UT across Solidity and Vyper. This alignment ensures that semantically equivalent vulnerability cues retain similar representations even when implemented with language-specific constructs. For example, improved alignment enables the classifier to correctly identify Reentrancy vulnerabilities despite differences such as the use of \texttt{MODIFIER\_CALL} in Solidity versus \texttt{INTERNAL\_CALL} in Vyper. These aligned representations allow the downstream classifier to generalize effectively, explaining the negligible performance drop when transferring from Solidity to Vyper.

\subsection{RQ3: Baselines Comparison}

We next compare the vulnerability detection performance of \toolvyper\ against several baselines. Existing approaches for detecting vulnerabilities in Vyper smart contracts rely primarily on traditional program analysis techniques, and no deep learning–based models currently exist for Vyper. To provide a comprehensive evaluation, we design three neural baselines adapted from state-of-the-art models. \MH{Additionally, we include an LLM-based baseline using direct prompting.} Below are the baselines included in our evaluation:

\begin{itemize}

\item Slither~\cite{feist2019slither}: A static analysis tool that extracts ASTs and detects vulnerabilities via manually crafted pattern matching

\item Mythril~\cite{sharma2022survey}: A dynamic analysis tool that applies symbolic execution to EVM bytecode generated from Vyper contracts.

\item Vyper-Trained Classification Network: A baseline where we directly train and test our classification network on Vyper, without any transferable knowledge learning. 

\item CodeT5+~\cite{wang2023codet5+}: A Transformer-based model pre-trained on a large multi-language corpus of source code.

\item GraphCodeBERT~\cite{guographcodebert}: A Transformer-based model pre-trained on multiple programming languages using both sequential context and data-flow information.

\item \MH{LLM Prompting: We provide vulnerability descriptions for each vulnerability type and prompt the LLM for detection. We select three LLM backends that demonstrate high performance on code: GPT-4.1~\cite{gpt41}, CodeLlama-7b~\cite{roziere2023code}, and Qwen-Coder-7b~\cite{bai2023qwen}.}  

\item \MH{Joint Training Baseline: To validate whether freezing is necessary for cross-language generalization, we evaluate a joint training variant where we do not freeze the multi-view encoders during Stage \circled{2}. Instead, we fine-tune the encoders and classifier end-to-end on the Solidity vulnerability dataset while concurrently applying the MMD alignment objective on unlabeled Solidity/Vyper batches. Specifically, we use the loss function $\mathcal{L}_{\text{Joint}}=\mathcal{L}_{\text{CLS}}+\lambda \mathcal{L}_{\text{MMD}}$ instead of original loss function $\mathcal{L}_{\text{MMD}}$. This baseline tests whether supervised gradients improve discriminative power at the cost of language invariance, compared to the frozen strategy.}  

\end{itemize}

For both CodeT5+ and GraphCodeBERT, we append a sigmoid classification layer on top of the encoder and fine-tune the model for the vulnerability detection task. 

For baseline comparison, the Vyper vulnerability detection dataset $D_{Vy-all}$ is split into two parts: 80\% for training (denoted as $D_{Vy-1}$), and 20\% for testing (denoted as $D_{Vy-2}$).

\vspace{3pt}
\noindent \textbf{Results.}
The results are presented in Table~\ref{tab:rq3}. We observe that \toolvyper\ consistently outperforms all baseline methods. Below, we provide a detailed analysis of these results.

\emph{\textbf{Comparison with Program Analysis-based Methods}.}
For Slither, which is based on static analysis techniques, consider reentrancy as an example. \toolvyper~achieves an \MH{FPR/FNR of 0.11/0.16 compared to Slither's 0.46/0.58}. This large performance gap arises from the \emph{pattern-matching limitation} in Slither's approach when applied to Vyper contracts. Slither relies heavily on manually crafted syntactic patterns, which fail in Vyper due to the language's different syntactic structures and programming paradigms. Our \toolvyper~abstracts away these syntactic differences, enabling pattern recognition at the semantic level.

For Mythril, which is based on symbolic execution, \toolvyper~still outperforms it. Mythril’s effectiveness is constrained by the path explosion problem, where the number of execution paths grows exponentially with program complexity. As a result, Mythril often fails to explore deep or complex contract behaviors and incurs significant computational overhead.


\emph{\textbf{Comparison with Deep-Learning-based Models}.}
Note that no deep learning–based models currently exist for Vyper, so we design three neural baselines.
The first is a \emph{Vyper-trained classification network}, where we directly train and test the classifier on Vyper contracts without any transferable-knowledge learning. As expected, this Vyper-only model performs worse than \toolvyper, primarily due to the \emph{data scarcity problem}: the available Vyper training set is too small to support training a robust detection model. Moreover, collecting a large number of vulnerable Vyper samples is difficult, which motivates our approach of reusing a well-trained Solidity model to detect vulnerabilities in Vyper contracts.

For the two state-of-the-art pretrained code models: GraphCodeBERT and CodeT5+. Both models are pretrained on large-scale multi-language code corpora and have demonstrated strong performance on various code understanding tasks. We attach a classification head to each model, fine-tune them on $D_{Vy-1}$, and evaluate them on $D_{Vy-2}$. GraphCodeBERT achieves \MH{FPR/FNR scores of 0.32/0.38, 0.35/0.42, and 0.33/0.39} for RE, WR, and UT, respectively, while CodeT5+ performs slightly better with scores of \MH{0.28/0.34, 0.32/0.38, and 0.30/0.36 for FPR/FNR}. In comparison, \toolvyper\ substantially outperforms both models.

\emph{\textbf{\MH{Comparison with LLM-based Models}}.}
\MH{We further compare~\toolvyper~with three LLM-based baselines (GPT-4.1, CodeLlama-7b, and Qwen-Coder-7b) using direct prompting for vulnerability detection. Specifically, we provide the full contract code together with a vulnerability description prompt and ask the model to determine whether the contract is vulnerable. As shown in Table~\ref{tab:rq3},~\toolvyper~consistently outperforms all LLM-based methods across RE, WR, and UT. For example, on RE detection,~\toolvyper~achieves an FPR/FNR of 0.11/0.16, significantly lower than GPT-4.1 (0.41/0.52), CodeLlama-7b (0.45/0.55), and Qwen-Coder-7b (0.39/0.50). This performance gap indicates that while LLMs possess general code understanding capabilities, simple prompting without explicit structural modeling or cross-language alignment is insufficient for accurate vulnerability detection in Vyper.}

\emph{\textbf{\MH{Comparison with Joint Training Baseline}}.}
\MH{Instead of freezing the multi-view encoders in Stage \circled{2}, 
we additionally evaluate a joint-training baseline that does not freeze encoder parameters. In this baseline, we fine-tune the encoders and classifier end-to-end on the labeled Solidity vulnerability dataset while simultaneously applying the MMD alignment objective on unlabeled Solidity/Vyper batches. The overall objective is $\mathcal{L}_{\text{Joint}}=\mathcal{L}_{\text{CLS}}+\lambda\mathcal{L}_{\text{MMD}}$, which tests whether freezing is essential for superior cross-language generalization. As shown in Table~\ref{tab:rq3}, the joint training only achieves FPR/FNR scores of 0.48/0.53, 0.68/0.78, 0.47/0.56 for RE, WR and UT, which is significantly higher than~\toolvyper. These results suggest that in the joint training paradigm, end-to-end supervised updates partially overwrite language-agnostic features learned in Stage \circled{1}. In contrast, freezing preserves the aligned representation space, enabling the Solidity-trained classifier to generalize with minimal degradation on Vyper.}

\subsection{RQ4: Hyperparameter Study}

To understand the impact of kernel composition on transferable knowledge learning effectiveness, we conduct a hyperparameter study focusing on $\alpha$ and $\gamma$ parameters in Equation~(\ref{equ:kernel}). We systematically vary $\alpha$ and $\gamma$ to evaluate their combined effect on vulnerability detection performance. For each parameter combination, we train~\toolvyper~using the same general dataset and Solidity vulnerability detection dataset, and evaluate on all three vulnerability types using the same Vyper test dataset.  We report the AUC score for each setting.  

\begin{figure}[t]
    \centering
    \includegraphics[width=0.99\columnwidth]{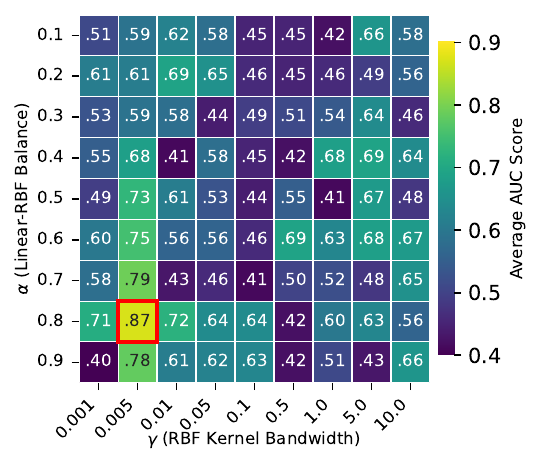}
    \aaf \aaf
    \caption{Heatmap showing the average AUC value when varying the hyperparameters $\alpha$ and $\gamma$.}
    \label{fig:heatmap}
    
\end{figure}

\vspace{3pt}
\noindent \textbf{Results.}
Figure~\ref{fig:heatmap} presents the average AUC scores across three vulnerability types for different hyperparameter combinations. We observe that the optimal combination is $\alpha=0.8$ and $\gamma=0.005$, achieving an average AUC of 0.87. This configuration can be explained in two aspects: (1) \emph{High linear component emphasis:} The optimal configuration heavily favors the linear kernel component ($\alpha=0.8$), suggesting that the domain gap between Solidity and Vyper is primarily characterized by linear transformations rather than complex nonlinear mappings. This indicates that while the two languages have syntactic differences, their semantic structures maintain largely linear relationships in the representation space. (2) \emph{Fine-grained RBF Bandwidth:} The very small $\gamma$ value ($0.005$) captures long-range semantic similarities while being robust to local syntactic variations. This bandwidth setting ensures that semantically similar code patterns across languages are mapped to nearby points in feature space. Larger $\gamma$ values (over 0.1) create overly narrow kernels that fragment the feature space and fail to generalize across language boundaries.

\vspace{3pt}
\noindent \textbf{Sensitivity Analysis.}
The performance landscape exhibits significant variability, indicating high sensitivity to parameter selection. (1) \emph{$\alpha$ sensitivity:} Performance shows a sharp peak at $\alpha = 0.8$, with substantial degradation at both higher ($\alpha = 0.9$) and lower ($\alpha \leq 0.7$) values. This suggests a critical threshold where linear alignment mechanisms become dominant for effective cross-language transfer. (2) \emph{$\gamma$ sensitivity} The optimal $\gamma = 0.005$ represents a narrow sweet spot. Both smaller ($\gamma = 0.001$) and larger ($\gamma \geq 0.01$) values show decreased performance, indicating that $\gamma$ must be precisely tuned to capture the appropriate scale of semantic similarities.

\subsection{RQ5: Ablation Study} \label{subsec:ablation}

\begin{table}[t]
\caption{Ablation Study Results.}
\aaf
\centering
\label{tab:ablation-result}
\resizebox{\columnwidth}{!}{
\begin{tabular}{l|cc|cc|cc}
\hline
                          & \multicolumn{2}{c|}{RE} & \multicolumn{2}{c|}{WR} & \multicolumn{2}{c}{UT} \\ \cline{2-7} 
                          & FPR        & FNR       & FPR        & FNR       & FPR        & FNR       \\ \hline \hline 
\toolvyper                & \multicolumn{1}{c|}{\textbf{0.11}}& \textbf{0.16} & \multicolumn{1}{c|}{\textbf{0.10}}& \textbf{0.15}& \multicolumn{1}{c|}{\textbf{0.12}}       &\textbf{ 0.17}      \\ \hline

\toolvyper~w/o seq.                  & \multicolumn{1}{c|}{0.42}       & 0.48      & \multicolumn{1}{c|}{0.29}       & 0.25      & \multicolumn{1}{c|}{0.31}       & 0.39      \\ \hline
\toolvyper~w/o hie.                  & \multicolumn{1}{c|}{0.28}       & 0.34      & \multicolumn{1}{c|}{0.55}       & 0.52      & \multicolumn{1}{c|}{0.28}       & 0.35      \\ \hline
\toolvyper~w/o both                  & \multicolumn{1}{c|}{0.45}       & 0.52      & \multicolumn{1}{c|}{0.58}       & 0.54      & \multicolumn{1}{c|}{0.46}       & 0.73      \\ \hline
\end{tabular}
}
\aaf
\end{table}

To isolate the contribution of each component in~\toolvyper, we conduct an ablation study that systematically removes the key elements of \toolvyper. Specifically, we consider the following four scenarios for comparison: 
\begin{itemize}
    \item \toolvyper: The complete framework with both sequential and hierarchical encoders, trained with MMD-based transferable knowledge learning.
    \item \toolvyper~without sequence encoder: This variant uses only the hierarchical encoder during both the unsupervised transferable knowledge learning stage and the vulnerability detection stage, without the sequential encoder.
    \item \toolvyper~without hierarchical encoder: This variant uses only the sequential encoder throughout the pipeline, without the hierarchical encoder. 
    \item \toolvyper~~without both sequence and hierarchical encoder (no transferable learning): This variant is trained directly on Solidity and tested on Vyper without learning the transferable knowledge.
\end{itemize}

For each configuration, we train the model using the same Solidity vulnerability detection dataset and evaluate on the Vyper test set across all three vulnerability types (RE, WR, UT). We report both FPR and FNR metrics on the vulnerability detection performance. The results are shown in Table~\ref{tab:ablation-result}. From the results, we have the following findings:

\vspace{3pt}
\noindent \textbf{Finding 1: Complementary Contributions of the Multi-View Encoders.}
From Table~\ref{tab:ablation-result}, we can see that~\toolvyper~with both hierarchical encoder and sequential encoders perform best. While remocostumed variable nameving either encoder results in a performance degradation for all the vulnerability types and metrics. Removing both hierarchical and sequential encoders leads to the most severe performance decrease. These results indicate that the sequential and hierarchical encoders provide \textbf{\emph{non-redundant}} views of SlithIR semantics, and that the full multi-view architecture is necessary to fully realize the benefits of transferable knowledge.

\vspace{3pt}
\noindent \textbf{Finding 2: Different Vulnerability Types Respond Differently to Component Removal.}
Another pattern revealed is that the three vulnerability categories respond differently to the removal of specific components, reflecting their inherent semantic structures and the type of invariances required for cross-language transfer.

For example, reentrancy attacks are fundamentally defined by a \textbf{\emph{temporal dependency pattern}} of: external call, state update and reentrant invocation path. Even minor deviations in SlithIR order can obscure this signal. Vyper’s aggressive inlining of modifier-like logic causes substantial operation-order drift, which the sequential encoder compensates for. Removing the sequential encoder increases FPR/FNR of RE from 0.11/0.16 to 0.42/0.48. While removing the hierarchical encoder only increases FPR/FNR from 0.11/0.16 to 0.28/0.34, which is relatively minimal. Thus, the RE detection requires \textbf{\emph{mostly temporal invariance}}, which only the \textbf{\emph{sequential encoder}} can provide. 

On the other hand, WR vulnerabilities rely on identifying flawed entropy sources buried inside \textbf{\emph{nested conditions}}, including multi-branch logic or deep AST regions. These patterns produce \textbf{\emph{long-range structural dependencies}} but are not highly sensitive to local instruction order. This is consistent with the results: Removing the hierarchical encoder makes the FPR/FNR of WR increase from 0.10/0.15 to 0.55/0.52, which is nearly identical to the worst scenario  (removing both encoders) of 0.58/0.54. So WR detection requires \textbf{\emph{structural invariance}}, which heavily depends on the \textbf{\emph{hierarchical encoder}}. For UT, an unchecked transfer combines a structural check and a sequential misuse of balance transfer. This hybrid nature explains why performance on UT drops moderately, but not catastrophically when either encoder is removed.

\subsection{RQ6: Few-shot Analysis} \label{subsec:few-shot}

To investigate the potential benefits of incorporating limited Vyper data during training, we conduct a few-shot analysis to examine how small amounts of labeled Vyper contracts influence \toolvyper’s performance. This analysis captures a practical scenario in which practitioners may have access to only a handful of labeled Vyper samples and wish to leverage them to further improve detection accuracy beyond the pure zero-shot transfer setting.

We evaluate fifteen different few-shot configurations by adding varying numbers and compositions of Vyper smart contracts to the training set, in addition to the original Solidity vulnerability detection data. The configurations span from 4-shot to 20-shot scenarios with three composition strategies: (1) \textbf{Safe-only additions}: Adding only safe Vyper smart contracts (4S, 8S, 12S, 16S, 20S), (2) \textbf{Vulnerable-only additions}: Adding only vulnerable Vyper smart contracts (4V, 8V, 12V, 16V, 20V), and (3) \textbf{Balanced additions}: Adding equal numbers of safe and vulnerable Vyper smart contracts (2S+2V, 4S+4V, 6S+6V, 8S+8V, 10S+10V). Note that the added Vyper smart contracts \emph{have no overlap} with the Vyper contracts for testing. For each configuration, we retrain the classification network while maintaining the same frozen multi-view encoder architecture from the unsupervised transferable knowledge learning stage. To ensure statistical robustness, we repeat each few-shot configuration with three different random samples and report the average performance.

\begin{table*}[t]
\centering
\caption{Few-shot analysis results (S = safe Vyper contracts; V = vulnerable Vyper contracts).}
\aaf
\label{tab:few-shot-analysis}
\resizebox{0.92\textwidth}{!}{
\setlength{\tabcolsep}{0.8mm}
\begin{tabular}{cc|c|ccc|ccc|ccc|ccc|ccc}
\hline
\multirow{2}{*}{} & \multirow{2}{*}{} & \textbf{0-shot} & \multicolumn{3}{c|}{\textbf{4-shot}} & \multicolumn{3}{c|}{\textbf{8-shot}} & \multicolumn{3}{c|}{\textbf{12-shot}} & \multicolumn{3}{c|}{\textbf{16-shot}} & \multicolumn{3}{c}{\textbf{20-shot}} \\
\cline{4-18}
& &(base)    & 4S   & 4V   & 2S+2V& 8S   & 8V   & 4S+4V& 12S  & 12V  & \textbf{6S+6V}& 16S  & 16V  & 8S+8V& 20S  & 20V  & 10S+10V \\ 
\hline\hline
\multicolumn{1}{c|}{\multirow{2}{*}{RE}} 
& FPR & 0.11 & 0.10 & 0.27 & 0.09 & 0.09 & 0.22 & 0.08 & \textbf{0.05} & 0.17 & 0.08 & 0.07 & 0.20 & 0.08 & 0.06 & 0.29 & 0.08 \\
\multicolumn{1}{c|}{}                    
& FNR & 0.16 & 0.25 & 0.13 & 0.12 & 0.23 & 0.11 & 0.10 & 0.18 & \textbf{0.07} & 0.08 & 0.29 & 0.10 & 0.07 & 0.28 & 0.09 & 0.08 \\
\hline

\multicolumn{1}{c|}{\multirow{2}{*}{WR}} 
& FPR & 0.10 & 0.09 & 0.28 & 0.08 & 0.08 & 0.22 & 0.07 & \textbf{0.05} & 0.19 & 0.06 & 0.09 & 0.28 & 0.07 & 0.08 & 0.31 & 0.07 \\
\multicolumn{1}{c|}{}                    
& FNR & 0.15 & 0.24 & 0.13 & 0.12 & 0.23 & 0.11 & 0.11 & 0.20 & \textbf{0.08} & 0.09 & 0.29 & 0.10 & 0.09 & 0.29 & 0.10 & 0.08 \\
\hline

\multicolumn{1}{c|}{\multirow{2}{*}{UT}} 
& FPR & 0.12 & 0.11 & 0.27 & 0.10 & 0.10 & 0.23 & 0.09 & \textbf{0.07} & 0.19 & 0.08 & 0.10 & 0.32 & 0.07 & 0.09 & 0.31 & 0.07 \\
\multicolumn{1}{c|}{}                    
& FNR & 0.17 & 0.25 & 0.14 & 0.13 & 0.24 & 0.12 & 0.11 & 0.18 & 0.08 & \textbf{0.06} & 0.30 & 0.11 & 0.08 & 0.27 & 0.10 & 0.08 \\
\hline

\end{tabular}
}
\end{table*}

\vspace{3pt}
\noindent \textbf{Results.}
Table~\ref{tab:few-shot-analysis} presents the few-shot analysis results across all three vulnerability types. We have the following findings: 

\begin{itemize}
    \item \textbf{Progressive Performance Gains}: \MH{Adding small amounts of balanced Vyper training data consistently improves performance across all vulnerability types.} The gains are most pronounced when moving from zero-shot to few-shot scenarios, with diminishing returns as more contracts are added.
    \item \textbf{Saturation Effects}: Performance improvements exhibit strong saturation effects beyond 16-20 examples, suggesting that~\toolvyper~can effectively leverage small amounts of target-domain data without requiring extensive labeled datasets. This saturation behavior validates our transferable knowledge learning approach: the frozen encoders already capture most cross-language semantic patterns, requiring only minimal fine-tuning through target-domain examples to bridge the remaining gap. The practical implication is significant since practitioners can achieve near-optimal performance by annotating just 12-16 carefully selected Vyper contracts. While further adding more Vyper samples may lead to overfitting.
    \item \textbf{Composition Sensitivity}: The composition of few-shot examples significantly impacts performance outcomes, with balanced datasets containing both safe and vulnerable contracts striking a good trade-off between FNR and FPR. This outcome is because a balanced combination can \textbf{\emph{promote inter-class discrimination}}, allowing encoder decision boundaries to be calibrated against both normal and risky samples. Additionally, balanced sampling minimizes feature space collapse towards a \textbf{\emph{single-class direction}}, leading to more stable classifier gradients in Stage \circled{2} fine-tuning. While including only vulnerable samples or safe samples will lead to a relatively high FPR or FNR rate.
\end{itemize}


\vspace{3pt}
\noindent \textbf{Shot Selection Strategy.}
In high-risk environments where omission of vulnerabilities is unacceptable, slightly biasing few-shot samples toward vulnerable contracts (e.g., 0S+8V) may further minimize the FNR rate, even though such configurations could introduce a moderate increase in FPR rate. For example, configurations such as 0S+8V in the 8-shot setting and 4S+8V in the 12-shot setting prioritize exposure to abnormal execution behaviors, encouraging the classification layer to adopt a more conservative decision boundary. This design shifts the model toward recognizing subtle vulnerability patterns rather than requiring high confidence from structural normality cues. As a result, the classifier becomes more strict on potential risk, effectively \emph{detecting more true vulnerabilities} at the cost of occasionally \emph{misclassifying safe contracts}. This pattern is aligned with practitioners’ intent in high-risk environments, such as DeFi protocols or DAO treasury management, where missing a vulnerability can cause irreversible financial impact, while falsely alerting a safe contract typically incurs only minor manual validation overhead.


\vspace{3pt}
\noindent \textbf{Cross-Vulnerability Transfer.}
We further explore whether few-shot examples for one vulnerability type can benefit the detection of other types. To investigate this, we conduct experiments where we add few-shot examples labeled for RE and measure the impact on WR and UT detection. Our results show limited cross-vulnerability transfer: adding 8 reentrancy-specific Vyper contracts improves reentrancy detection by 0.05 AUC but provides only 0.01 improvement for WR and UT. This suggests that vulnerability-specific patterns are largely orthogonal in the feature space, and practitioners should allocate annotation effort proportionally across the vulnerability types they care about rather than focusing on a single type.

\subsection{RQ7: Runtime Overhead Analysis}

To assess the practical applicability of~\toolvyper, we conduct a runtime overhead analysis comparing our approach against traditional static and dynamic methods. We measure the end-to-end analysis time for each tool on the Vyper vulnerability detection dataset, which contains contracts ranging from 50 to 800 lines of code. 
For each tool, we measure: (1) preprocessing time (including compilation and IR generation where applicable), (2) core analysis time, and (3) total wall-clock time per contract. We exclude dataset loading and initialization overhead, focusing solely on per-contract analysis time.

\MH{To reduce noise, we select different seeds and run~\toolvyper~and Mythril five times for each contract. Then we report the mean runtime with 95\% confidence intervals (CI). Slither is a static deterministic tool so we only run it once. This provides a statistically supported estimate of the typical runtime and quantifies uncertainty due to run-to-run variation.}

\begin{table}[t]
\caption{\MH{Runtime overhead: average time (s) per contract. We report the mean $\pm$ CI over five runs.}}
\aaf
\label{tab:runtime-overhead}
\centering
\resizebox{\columnwidth}{!}{
\begin{tabular}{l|c|c|c}
\hline
Contract Size & \MH{\toolvyper} & \MH{Mythril} & Slither \\ \hline \hline
Small (20-100 lines) & \MH{0.28$\pm$ 0.16}& \MH{24.3 $\pm$ 21.1} & 0.15 \\  \hline
Medium (101-200 lines) & \MH{0.42$\pm$ 0.13}& \MH{87.0 $\pm$ 43.6}& 0.21 \\ \hline
Large ($\ge$ 201 lines) & \MH{0.65$\pm$ 0.21}& \MH{303 $\pm$ 264.4}& 0.31 \\ \hline
\end{tabular}
}
\aaf
\end{table}

\vspace{3pt}
\noindent \textbf{Results.}
Table~\ref{tab:runtime-overhead} presents the runtime overhead comparison across different contract size categories.~\toolvyper\ demonstrates superior efficiency compared to dynamic analysis while maintaining competitive performance with static methods. For small contracts (20-100 lines),~\toolvyper\ achieves an average analysis time of 0.28 seconds, slightly higher than Slither's 0.15 seconds but orders of magnitude faster than Mythril's 24.3 seconds. For medium-sized contracts (101-200 lines),~\toolvyper's analysis time increases sublinearly to 0.42 seconds, while Mythril's time explodes to 87.0 seconds due to path explosion in symbolic execution. For large contracts (over 201 lines), the efficiency gap widens further:~\toolvyper\ requires only 0.65 seconds on average, compared to Mythril's over 300 seconds. Notably, Slither still maintains the fastest analysis time for all scales. But as shown in Table~\ref{tab:rq3}, this speed comes at the cost of significantly lower detection accuracy (FPR of 0.45-0.48 compared to~\toolvyper's around 0.11).

\vspace{3pt}
\noindent \textbf{Analysis.}
The efficiency of~\toolvyper~stems from the architectural advantage of one-pass encoding: Unlike Mythril's iterative symbolic execution,~\toolvyper\ performs a single forward pass through the multi-view encoders and classification network. These results demonstrate that~\toolvyper~achieves an optimal balance between detection accuracy and computational efficiency, making it suitable for both on-demand analysis and large-scale automated scanning pipelines.

\section{\MH{Discussion on Generalization}}
\MH{Considering that Solidity is the most common language in blockchain, we have developed techniques to transfer knowledge from Solidity to Vyper. To validate the effectiveness of the techniques, we conducted an evaluation that tested the transferability from Solidity to Vyper for the downstream vulnerability detection task. In theory, the techniques could apply generally. However, it is important to note that our results are specific to the evaluated scenarios. In order to ascertain the effectiveness of this approach across other downstream tasks and other smart contract languages, further investigation and comprehensive testing are needed.}

\section{Conclusion}

In this work, we introduced \toolvyper, the first framework enabling deep learning–based vulnerability detection for Vyper smart contracts without requiring any labeled Vyper training data. The core insight is that SlithIR, a language-agnostic intermediate representation, bridges Solidity and Vyper by abstracting away syntactic differences while preserving essential semantics. Leveraging this, we designed a three-stage framework that integrates unsupervised transferable knowledge learning, supervised vulnerability detection on Solidity, and zero-shot testing on Vyper. Experiments across key vulnerability types, including reentrancy, weak randomness, and unchecked transfer, show that \toolvyper\ achieves strong zero-shot performance on Vyper and significantly outperforms prior methods lacking cross-language transfer.


\bibliographystyle{IEEEtran}
\bibliography{custom}

\newpage

\end{document}